\documentstyle[aps,prc,epsfig]{revtex}
\newcommand{\beq}{\begin{equation}}
\newcommand{\eeq}{\end{equation}}
\newcommand{\bea}{\begin{eqnarray}}
\newcommand{\eea}{\end{eqnarray}}
\newcommand{\bce}{\begin{center}}
\newcommand{\ece}{\end{center}}
\newcommand{\eg}{{\it e.g.}}
\newcommand{\ie}{{\it i.e.}}
\newcommand{\etal}{{\it et al.}}
\def\lsim{\mathrel{\rlap{\lower4pt\hbox{\hskip1pt$\sim$}}
    \raise1pt\hbox{$<$}}}         
\def\gsim{\mathrel{\rlap{\lower4pt\hbox{\hskip1pt$\sim$}}
    \raise1pt\hbox{$>$}}}         
\begin{document}
\twocolumn[\hsize\textwidth\columnwidth\hsize\csname @twocolumnfalse\endcsname
%
%
\title{Resolving the Antibaryon-Production Puzzle  
 in High-Energy Heavy-Ion Collisions}
 
\author{Ralf Rapp and Edward V. Shuryak}
 
\address
{Department of Physics and Astronomy, State University of New York, 
    Stony Brook, NY 11794-3800, U.S.A.}

\date{\today} 

\maketitle
 
\begin{abstract}
We argue that the observed antiproton production in heavy-ion
collisions at CERN-SpS energies can be understood if 
(contrary to most sequential scattering  approaches)
the backward direction in the process  
$p\bar p \leftrightarrow \bar{n}\pi$  (with $\bar{n}$=5-6)
is consistently accounted for within a thermal framework.   
Employing the standard picture of subsequent
chemical and thermal freezeout, which induces an over-saturation of pion 
number with associated chemical potentials of  $\mu_\pi\simeq$~60-80~MeV,  
enhances the backward reaction substantially. The resulting rates
and corresponding cross sections turn out to be large enough to 
maintain the abundance of antiprotons at chemical freezeout  
until the decoupling temperature, in accord with the measured 
$\bar{p}/p$ ratio in Pb(158AGeV)+Pb collisions.
\end{abstract} 
\vspace{0.58cm}
]
\begin{narrowtext}
\newpage

Over the last decade remarkable progress has been made in the 
understanding of the dynamics of strong interactions probed through 
(ultra-) relativistic heavy-ion collisions.
Although the main challenge of an unambiguous identification of the QCD
phase transition to the Quark-Gluon Plasma (QGP) persists,
we have greatly advanced our knowledge on properties of highly excited
hadronic matter close to the expected phase boundary. A variety of 
collective phenomena has been observed indicating that the produced  
systems have indeed reached macroscopically large sizes, justifying the
use of equilibrium techniques such as thermo- and hydrodynamics.
  
One of the important results that will be used below is that the 
final-state hadron abundances, including antibaryons, can be rather 
accurately characterized by the so-called {\em chemical} freezeout 
stage~\cite{pbm96-99} with a common temperature $T_{ch}$ and baryon 
chemical potential $\mu_{B}^{ch}$, the specific values depending 
on collision energy (note that a precise description of all hadron
species requires corrections to an ideal gas ensemble, {\eg},  
excluded (eigen-) volumes~\cite{pbm96-99,YG99} to mimic short-range
repulsions, or 'strangeness suppression' factors~\cite{YG99}.
Such corrections are not important for our subsequent analysis 
and will be neglected. For a contrasting view of hadron production 
in heavy-ion collisions, see, {\eg}, ref.~\cite{RL99}). 

At SpS energies, the chemical freezeout is clearly distinct
from the {\em thermal} one (with an associated temperature 
$T_{th}<T_{ch}$), from where on the particles stream freely to 
the detectors. This follows from the kinetics of hadronic 
reactions~\cite{kinetics,HS98}, \ie, a significant difference 
between elastic and inelastic collision 
rates at low relative energies, and has been confirmed by 
several experimental evidences (see, \eg, ref.~\cite{Stock}).  
In central Pb+Pb collisions, \eg,  a nucleon at midrapidity
is elastically rescattered on average about 10-15 times, but less 
than once inelastically~\cite{rqmd95}.  
Also, as shown in \cite{HS98}, most of the collective flow effects 
at SpS are generated {\em in between}
the two freezeouts, and their observation leaves no doubt
about the existence of such an  intermediate stage.

Another consequence is that abundances of secondary mesons
(pions, kaons, etc.) are not subject to significant changes when 
the system evolves from $T_{ch}$ to $T_{th}$. Again, this  
is conceivable as the rates for number changing reactions are 
too small to maintain chemical equilibrium on the time scales 
of the hadronic fireball lifetime,  
$t\simeq 10$~fm/c. Since, on the other hand,  elastic scattering 
(\eg,  $\pi\pi\to\rho\to \pi\pi$) is still effective, 
thermal equilibrium is approximately maintained.
In a statistical mechanics language such a scenario entails additional
conservation laws, which can be implemented via effective  
(pion-, kaon-, etc.) chemical potentials to guarantee fixed     
particle numbers. For pions (kaons) 
typical values of $\mu_\pi=$~60-80~MeV ($\mu_K=$~100-130~MeV)
are reached at $T_{th}=$~110-120~MeV, see \cite{kinetics,HS98,RW99}.
The ensuing 'over-saturation' of the pion phase space
is to play the key role in what follows.

The situation for antibaryons is different from mesonic secondaries
as the pertinent  {\em inelastic} (or annihilation) cross section is 
{\em not} small. At the relevant (thermal) energies
in the comoving frame of collective expansion, 
$\sqrt{s}= 2(m_N+E_{N}^{th})\simeq 2.3$~GeV, one has   
$\sigma_{p\bar p\to n\pi} \simeq 50$~mb.  
Taking an average baryon density of $\varrho_B=0.75\varrho_0$ 
($\varrho_0=0.16$~fm$^{-3}$) around $T=150$~MeV in the 
course of the hadronic evolution and a typical (anti-) proton velocity 
of $v_{th}=p/E_{tot}=0.56c$ (from $E_{p}^{th}=(3/2) kT \simeq 225$~MeV), 
we obtain for the chemical equilibration time scale
\beq
\tau_{ch}=\frac{1}{\sigma_{ann} \ \varrho_B \ v_{th}}\simeq 3~{\rm fm/c} \ .  
\eeq
This is well below the fireball lifetime in the pure hadronic phase
of $\tau_{had}\simeq 7$~fm/c~\cite{HS98,RW99}, cf. Fig.~\ref{fig_tau};
 in other words, only a fraction
of $\exp[-\tau_{had}/\tau_{ch}] \simeq 10$~\% does {\em not} rescatter 
towards thermal freezeout. Thus, in spite of the good agreement of the
final $\bar p$ yields with the standard chemical freezeout predictions, 
antibaryon production ought to be reconsidered.

Naively one might expect most of the antiprotons to be annihilated, and 
various transport calculations (\eg, ARC~\cite{ARC} and UrQMD~\cite{Blei00})
have indeed  been unable to account for the measured number, falling short
by  significant factors. Consequently, speculations have been raised  
that the puzzle could be resolved by certain in-medium effects leading to 
either a reduction of the annihilation rate~\cite{ARC}
or an enhanced production~\cite{Blei00}. 

\begin{figure}
\epsfig{file=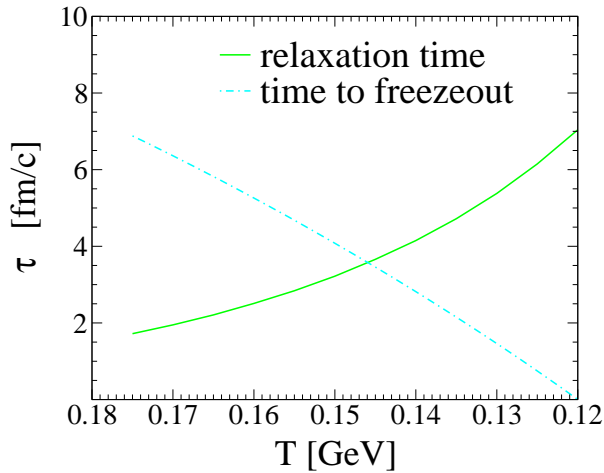,width=6.8cm,angle=-90}
\caption{Chemical relaxation time (full line) for antiprotons in central 
$Pb+Pb$ collisions at SpS energies employing an annihilation cross section 
of $\sigma_{ann}$=(40-50)~mb and a baryon-density profile as obtained
within an isentropically expanding fireball model~\protect\cite{RW99}.
The dashed-dotted curve indicates the remaining hadronic fireball lifetime 
till thermal freezeout.}
\label{fig_tau}
\end{figure}

A generic problem with transport/cascade simulations is that, although 
multi-body resonance decays are included, the inverse reactions are not. 
As is well-known, this violates detailed balance 
and, in principle, prevents the simulations from reaching the 
proper thermodynamic limit~\cite{Belk98}.  
The discussions are typically focused on three-body reactions;
in the present context it is even 5- or 6-pion collisions which 
are most relevant for producing baryon-antibaryon pairs. 
A widespread belief is that those reactions have insignificant   
rates under the conditions probed in the hadronic stages of heavy-ion
collisions.  However, as we will argue below, this is not  
the case; inverse reactions have to be addressed and are, in fact,  
capable of explaining the observed antiproton yield (although the $\bar p$
enhancement in nucleus-nucleus over $p$-$p$ collisions thus looses 
its proposed direct relation to QGP formation, the latter
is by no means excluded by the subsequent arguments).

The expression for the thermal reaction rate for the process 
$p\bar p \leftrightarrow {n}\pi$ can be written as (see also 
Ref.~\cite{KMR86})
\bea 
{\cal R}_{th} = \int 
d^3\tilde k_p~d^3\tilde k_{\bar p}~d^3\tilde k_{\pi_1} \cdots 
d^3\tilde k_{\pi_n} \ \delta^{(4)}(K_{tot}) \ |{\cal M}_n|^2  
\nonumber\\
\times \{ z_p~z_{\bar p} \ \exp[-\frac{E_p+E_{\bar p}}{T}] -
z_\pi^{ n} \ 
\exp[-\sum\limits_{i=1}^{ n} \frac{\omega_{\pi,i}}{T}]~\} \ , 
\label{rate}
\eea 
where ${\cal M}_n$ denotes the invariant scattering matrix element
(which, of course, is identical for the back- and forward direction 
due to time-reversal invariance of strong interactions), 
$d^3\tilde k_x= d^3k_x/(2\pi)^3$ the phase space integrations and
$K_{tot}=k_p+k_{\bar p}-k_{\pi_1}-\dots-k_{\pi_n}$ the total four-momentum;   
$z_x={\rm e}^{\mu_x/T}$ are the fugacities of particle species $x$
(in Boltzmann approximation), and $n$ is the number of pions  
produced in a $\bar pp$ annihilation at a given energy (to be discussed 
in more detail below). 
For a nonvanishing net nucleon number and in
chemical equilibrium, one has  $\mu_N>0$, $\mu_{\bar N}=-\mu_N$  and 
$\mu_\pi=0$. Under SpS conditions, typical values at chemical 
freezeout are~\cite{pbm96-99} $(T_{ch},\mu_{N}^{ch})\simeq(170,260)$~MeV, 
which results in  an antiproton-to-proton ratio of
\bea
\frac{\bar p}{p} & \propto & 
\frac{\exp[-(E_{\bar p}+\mu_N)/T]}{\exp[-(E_p-\mu_N)/T]}  
\nonumber\\
  &=& \exp[-2\mu_N/T] \ = \ 4.7\% \ , 
\eea
consistent with the experimentally measured value of 
$(5.5\pm 1)$~\%~\cite{na44,na49}.
Note also that at chemical freezout $z_p  z_{\bar p}=1$, $z_\pi=1$,
and the forward and backward rates in eq.~(\ref{rate}) are 
simply equal (the Boltzmann factors in both terms contain the 
same total energy). This implies, \eg, that cascade simulations 
ignoring the back-reaction at comparable pion densities 
($\varrho_\pi\simeq 0.2-0.25$~fm$^{-3}$) cannot give a proper account
of the antiproton production.  

The evolution of the system towards thermal freezeout can be constructed 
using a thermal fireball model which, based on isentropic expansion,
leads to $(T_{th},\mu_{N}^{th})\simeq(120,415)$~MeV. At first sight, 
this gives a $\bar p/p$ ratio of $\exp[-2\mu_N/T]\simeq 0.1$~\%,
a factor of $\sim$50 below the experimental result. 
However, this estimate is lacking an important ingredient.
A correct statistical treatment including pion-number 
conservation forces us to introduce individual
chemical potentials/fugacities for anti-/protons and pions. 
Insisting on equilibrium for the reaction in question, 
the following relation holds: 
\beq
z_{\bar p} = (z_p)^{-1} {\langle z_\pi^n\rangle} \ .   
\label{zpbar}
\eeq
Since antibaryons constitute only a small fraction of the produced 
secondaries, they have an insignificant impact on the evolution of
nucleon- and pion-chemical potentials. Therefore, eq.~(\ref{zpbar})
provides an estimate of the antiproton fugacity at thermal freezeout.   
The crucial point here is that pion over-saturation will generate  
a strong enhancement of the back-reaction $n\pi\to p\bar p$.

For a more quantitative assessment  
it is important to have a rather accurate determination of the 
pion multiplicity distributions  in  $p\bar p$ annihilation.
A nice discussion of its systematics has been given in 
ref.~\cite{Dov92}, which we will follow here to  
incorporate the empirical knowledge.  
For $\bar pp$ annihilation at rest ($\sqrt{s}=2m_N$), the data
can be represented by a least-square fit to a Gaussian probability 
distribution (normalized to one),
\beq
P(n)=\frac{1}{\sqrt{2\pi} \sigma} \ \exp[-(n-\langle n\rangle)^2/2\sigma^2] \ ,
\eeq 
with a mean $\langle n\rangle =5.02$ and width $\sigma=0.90$ (see also
ref.~\cite{BH95}). 
At larger $c.m.$-energies both the average number and width increase. For 
the former, linear~\cite{Greg76} and logarithmic~\cite{Sten79} 
dependencies have been proposed, 
\bea
\langle n \rangle &=& c_1 + c_2 \ s^{0.5}  
\nonumber\\
\langle n \rangle &=& \tilde{c}_1 + \tilde{c}_2 \ \ln(s) \     
\label{nav}
\eea
($s$ in [GeV$^2$]), which both reproduce the measured multiplicities
up to at least $\sqrt{s}=5$~GeV using the parameters
$c_1=2.6\pm0.5$, $c_2=(1.3\pm0.2)$~GeV$^{-1}$ and 
$\tilde{c}_1=2.65$, $\tilde{c}_2=1.78$,
respectively. For the latter, the energy dependence of the width has also
been given as~\cite{Sten79}
\beq
\sigma^2=0.174 \ \langle n \rangle \ s^{0.2} \ .  
\label{sig}
\eeq
For our application in a thermal environment at $T=150$~MeV 
(implying $\sqrt{s}=2.33$~GeV) 
we fix  $\langle n \rangle=5.65$ (in accordance with eq.~(\ref{nav})) 
together with a 10\% increase in $\sigma$ (as suggested by eq.~(\ref{sig})) 
to extract discrete weights $w_n=P(n;\langle n\rangle,\sigma)$. 
The averaged pion-fugacity enhancement factor then follows as
\beq
\langle z_\pi^{ n}\rangle=
\sum\limits_{n=2}^{n_{max}} w_n \ \exp[n \mu_\pi/T] \ ,  
\eeq
where $n_{max}=9$ for any practical purpose. Inserting now thermal
freezeout values $T_{th}=120$~MeV and $\mu_{\pi}^{th}\simeq 65$~MeV
(as arising in a thermal fireball model~\cite{RW99}), 
yields $\langle z_\pi^n\rangle =25$. This entails a large enhancement 
of the antiproton-to-proton ratio, from 0.1\% to 2.5\%.
In fact, owing to the high power of the pion fugacity,
slightly larger chemical potentials of $\mu_\pi=$~75-80~MeV
result in an enhancement factor of 42-54, rendering the pertinent 
$\bar p/p$-ratio in line with the observed (chemical freezeout) 
value, cf. Fig.~\ref{fig_ratio}.  
\begin{figure}
\epsfig{file=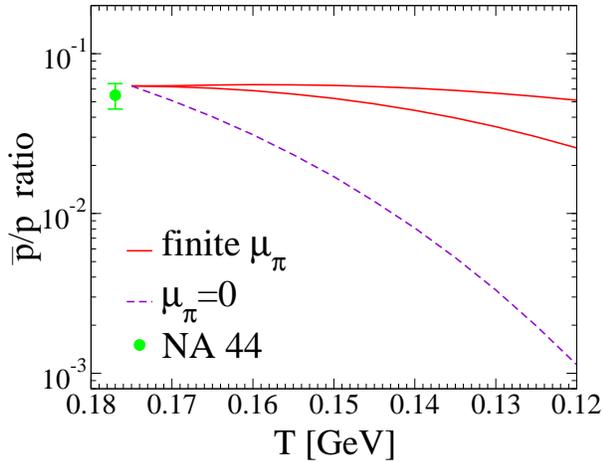,width=6.8cm,angle=-90}
\caption{Antiproton-to-proton ratio as a function of (decreasing) temperature
in an isentropically expanding fireball. The dashed curve represents the 
naive ratio, $\exp[-2\mu_N/T]$, whereas the full curves are for finite 
pion chemical potentials indicating  uncertainties as discussed in the
text. The experimenatl data point is from Ref.~\protect\cite{na44}.}
\label{fig_ratio}
\end{figure}

Such slightly increased  values for the pion chemical potential close 
to thermal freezeout can indeed be easily argued for. Within the 
thermal fireball model of ref.~\cite{RW99} elastic $\pi N \to B$ 
scattering ($B$: baryonic resonances up to $m_B\simeq 1.7$~GeV) was 
assumed to be frequent enough to maintain (relative) chemical 
equilibrium for the occupation of the excited baryonic states. 
However, with typical corresponding cross sections of 
$\sigma_{\pi N\to B}\simeq$~15-30~mb~\cite{pdg00}, this might 
not be fully justified anymore for the last few fm/c prior
to thermal freezeout. Consequently, a larger fraction of the pion number
resides in explicit pionic degrees of freedom rather than in excited 
resonances, which translates into an effectively larger $\mu_\pi$.

Let us finally comment on implications  of our observations
for RHIC. Close to the expected chemical freezeout the pion density is 
very similar to SpS conditions. Thus the rate of producing antiprotons 
through multi-pion annihilation per unit time and volume is essentially the 
same in both cases. The crucial difference is, however, that 
the {\em total} density of antiprotons is much larger around 
midrapidity at RHIC due to substantially smaller baryon chemical 
potentials. 
More quantitatively, using typical thermal model estimates~\cite{qm99}  
with $\varrho_B^{tot}\simeq 0.2\varrho_0$ shortly after chemical
freezeout (further reduced thereafter), one obtains
$\tau_{ch}^{RHIC}\simeq 11$~fm/c. With the lifetime
of the hadronic phase at RHIC being comparable to that at SpS energies, 
chemical equilibrium in the $p\bar p\leftrightarrow n\pi$ reaction 
cannot be maintained until thermal freezeout (also, the emerging 
pion oversaturation is less pronounced in a baryon-poor regime). 
The observed antiprotons 
at RHIC should therefore  mostly originate from earlier stages, 
corresponding to the standard hadro-chemical freezeout in the 
vicinity of the phase boundary. Nevertheless, our time scale estimate
indicates that even under RHIC conditions, antibaryon annihilation
will be partially compensated by the inverse reactions.

To summarize, we have analyzed the $\bar p /p$-ratio at SpS energies 
employing a thermal approach. So far this observable has been  
difficult to understand within, \eg, 
transport models  which only included the annihilation
channel, causing doubts whether the latter is actually active, 
or unconventional mechanisms for enhanced production need to be invoked.  
We have shown, however, that the 'puzzle' can be resolved 
in a rather standard statistical-mechanics framework  upon inclusion 
of the inverse process of multipion scattering    
into $\bar p p$ pairs, which can be supported until thermal
freezeout.  Our main ingredient was that effective
pion-number conservation generates pion over-saturation at the later
stages of a heavy-ion collision, as described by the build-up of 
appreciable pion chemical potentials. Raised to a large power ($n\sim 6$) 
the corresponding pion fugacities sustain a high antibaryon fraction, 
thus counter-balancing the loss from $B\bar B$ annihilation.
This mechanism also complies with the measured 
centrality dependence being essentially constant, as to be expected 
from a hadro-chemistry varying  little with impact parameter (for 
sufficiently peripheral collisions the applicability of thermal model 
analyses, of course, ceases and $\bar p$ production, normalized
to the number of participant nucleons,  approaches its
value in $p$-$p$ collisions, which lies about 30\% below the one 
in central nucleus-nucleus reactions~\cite{na49}).   

Finally we should note again that our findings are not contradictory
to an earlier QGP formation. They rather 
give further support to the equilibrium concept of subsequent
chemical and thermal freezeout stages, which   
has already proven successful in the explanation of a variety of  
observables, such as hadron abundances, collective flow, HBT radii
and electromagnetic radiation.  
 
\vskip0.2cm
 
This work was supported by the U.S. Department of Energy 
under Grant No. DE-FG02-88ER40388.

\end{narrowtext}
\end{document}